# Efficient and controlled domain wall nucleation for magnetic shift registers

Oscar Alejos[1], Víctor Raposo[2], Luis Sánchez-Tejerina[1], and Eduardo Martínez[2]*

[1] Dpto. Electricidad y Electrónica, Facultad de Ciencias, University of Valladolid, E-47011 Valladolid, Spain

[2] Dpto. Física Aplicada. University of Salamanca, Plaza de los Caídos s/n, E-38008, Salamanca, Spain

* Corresponding author: edumartinez@usal.es

**Ultrathin ferromagnetic strips with high perpendicular anisotropy have been proposed for the development of memory devices where the information is coded in tiny domains separated by domain walls. The design of practical devices requires creating, manipulating and detecting domain walls in ferromagnetic strips. Recent observations have shown highly efficient current-driven domain wall dynamics in multilayers lacking structural symmetry, where the walls adopt a chiral structure and can be driven at high velocities. However, putting such a device into practice requires the continuous and synchronous injection of domain walls as the first step. Here, we propose and demonstrate an efficient and simple scheme for nucleating domain walls using the symmetry of the spin orbit torques. Trains of short sub-nanosecond current pulses are injected in a double bit line to generate a localized longitudinal Oersted field in the ferromagnetic strip. Simultaneously, other current pulses are injected through the heavy metal under the ferromagnetic strip. Notably, the Slonczewski-like spin orbit torque assisted by the Oersted field allows the controlled injection of a series of domain walls, giving rise to a controlled manner for writing binary information and, consequently, to the design of a simple and efficient domain wall shift register.**

**Introduction**

A storage scheme based on racetrack memories[1,2,3], where the information is coded in magnetic domains (MDs) separated by adjacent domain walls (DWs), appears to be a promising alternative to hard disk drives for use as high-density storage devices. The initially suggested racetrack memory was based on in-plane magnetized systems in which head-to-head and tail-to-tail walls were driven by a spin current generated by the spin-transfer torque (STT)[4,5] mechanism. However, current research interest has been focused on ultrathin ferromagnetic (FM) strips sandwiched between a nonmagnetic heavy metal (HM) and an insulator; the resultant devices show a high perpendicular magnetocrystalline anisotropy (PMA) and exhibit anomalously efficient current-induced DW motion[6,7,8,9,10,11]. Upon the application of a current through the HM, the Slonczewski-like spin orbit torque (SL-SOT) resulting from the spin Hall effect (SHE)[12,13,14,15] generates sufficiently large spin currents to drive DWs in the FM layer if these walls adopt a chiral Néel configuration[7,8,15]. The Dzyaloshinskii-Moriya interaction (DMI)[16,17,18,19,20] at the interface between the HM layer and the FM layer enables the formation of such chiral Néel walls. In many cases, the spin Hall angle of the HM layer, together with the strength of the DMI, define the efficiency of the current-induced motion of chiral domain walls[7,11,20]. Narrower DW widths and low threshold current densities required for DW propagation and high velocities make these asymmetric PMA stacks a promising platform for solid-state magnetic devices based on electrically manipulated DWs[1,2,3].

Putting these DW-based devices into practice requires the controllable generation, manipulation and detection of individual domains separated by DWs for data writing, transmission and readout operations. Reading out the information encoded in a series of *up* ($m_z \approx +1$) and *down* ($m_z \approx -1$) magnetic domains can usually be accomplished using a magnetic tunnel junction localized at some reading point along the multilayer: with an out-of-plane pinned layer provided as a reference, the presence of an *up* or *down* domain is detected in the form of electrical resistance. The shifting or transmission of the information coded in the FM strip is efficiently performed by injecting current pulses along the HM ($\vec{J}_{HM}(t) = J_{HM}(t)\vec{u}_x$) via the spin Hall effect[7,8,9]. Current-driven DW velocities can reach $v \approx 350$ m/s under current pulses of $J_{HM} \approx 3$ TA/m² (Ref. 6). These characteristics, which make DW-based devices competitive with existing semiconductor and magnetic recording devices, have fuelled interest in these systems over the past decade. However, one of the main challenges that needs to be addressed in the development of such DW-based devices is the controlled nucleation of highly packed DWs.

The conventional procedure to nucleate DWs consists of the injection of an electrical current pulse along an adjacent conducting wire (bit line) orthogonal to the longitudinal axis ($x$) of the FM strip[20,21,22,23]. By starting from a uniform magnetic state ($m_z \approx \pm 1$), the Oersted field ($\vec{B}_{Oe} = \mu_0 \vec{H}_{Oe}$) generated by the current along the nucleation bit line ($I_L$)[7,23] locally reverses the magnetization in the FM ($m_z \approx \mp 1$), resulting in the nucleation of two DWs. In most of the related experimental studies, this procedure has been used as a preliminary step to evaluate the subsequent DW dynamics under current pulses injected along the HM/FM bilayer[7,23]. Typically, nucleation pulses along the bit line require amplitudes $I_L \approx$ 100 mA[23], and, as will be shown here, the perpendicular component of the generated Oersted field can reach substantial values ($B_{Oe,z} = \mu_0 H_{Oe,z} \approx$ 100 mT) even far from the nucleation line. Such a high field can also disturb the magnetic state of other DWs already present in the FM strip, eventually resulting in the destruction of the already coded information. Therefore, a need exists for a confined nucleation mechanism that does not perturb the magnetic state of other existing domains and domain walls already placed along the FM strip. An improvement of this nucleation procedure has recently been proposed by Zhang et al.[24], who designed a Π-shaped stripline to generate a strong localized out-of-plane field and to nucleate a domain wall with less energy than that required for the conventional single injection line.

Another nucleation scheme based on the STT has recently been proposed[25]. In this case, the local nucleation can be achieved in a region irradiated with ions to locally reduce or even cancel the PMA. In this case, the magnetization is essentially in-plane, forming a 90° magnetization boundary to the adjacent ferromagnetic strip with strong PMA. Due to the non-uniform magnetization at the transition between the in-plane and the out-of-plane magnetized areas, the conventional STT mechanism can promote the DW nucleation under current pulses directly flowing along the FM strip. However, although this scheme avoids the use of an external magnetic field, it relies on the STT mechanism, which has been shown to be negligible for most of the asymmetric PMA multilayers in which homochiral DWs are efficiently driven by the SHE[26,27]. A more recent experiment has shown that a single current pulse with the optimal length injected through the HM can also create a large number of closely spaced DWs[28]. However, local control of the DW nucleation remains elusive.

To address these drawbacks, here, we propose a simple and efficient method to nucleate homochiral DWs along an HM/FM multilayer with high PMA. This method eliminates the out-of-plane component of the nucleating Oersted field ($B_{Oe,z} \approx 0$) and uses the symmetry of the Slonczewski-like spin orbit torque to not only drive trains of DWs along the FM strip but also efficiently nucleate them under the presence of a localized in-plane field. Our results constitute a major step towards developing novel and efficient DW-based spintronic devices.

**Results**

**DW nucleation by a single bit line**

We first study the conditions needed to nucleate a pair of DWs by injecting a current pulse along a single bit line. A schematic of the considered geometry is shown in Fig. 1(a). The FM strip dimensions, as defined in Fig. 1(a), are $w \times t_{FM} = 192 \times 0.8$ nm$^2$. The cross section of the bit line is $w_L \times t_L = 200 \times 50$ nm$^2$, and its centre is located at $(x_L, z_L) = (-1.5$ μm$, t_d + \frac{t_L}{2})$ from the centre of the FM layer (0,0); $t_{FM}$ represents the thickness of the FM layer, and $t_d = \frac{t_{FM}}{2} + 3$ nm defines the vertical distance from the bottom edge of the bit line to the centre of the FM layer ($z = 0$). The Oersted field ($\vec{B}_{Oe}$) generated by the current $I_L$ passing through the bit line is computed numerically (by solving the Poisson equation) and/or analytically (using Biot-Savart's law), both of which lead to identical results[29] (see the Methods section).

The longitudinal ($B_{Oe,x}$) and perpendicular ($B_{Oe,z}$) components of the Oersted field ($\vec{B}_{Oe} = \mu_0 \vec{H}_{Oe}$) in the FM layer as a function of $x$ are shown in Figs. 1(b) and 1(c) for three different currents $I_L$. As expected, the magnitude of the Oersted field increases with $I_L$. Its longitudinal component $B_{Oe,x}$ is almost confined to the bit line width (see Fig. 1(b), $w_L = 200$ nm). The perpendicular component ($B_{Oe,z}$) achieves its maximum value at each edge of the bit line ($x = x_L \pm \frac{w_L}{2}$), and its magnitude decreases as $|B_{Oe,z}| \sim \frac{1}{|x-x_L|}$ for $|x - x_L| > \frac{w_L}{2}$.

Importantly, $|B_{Oe,z}|$ takes significant values even at points far from the nucleation point. Note that $|B_{Oe,z}| \approx 50$ mT at a distance $|x - x_L| \approx 800$ nm from the bit line placed at $x_L$. Such a large out-of-plane field, which is needed to nucleate DWs under the bit line, should also influence other DWs already present in the FM layer, and this effect should be taken into account in the design of DW-based devices.

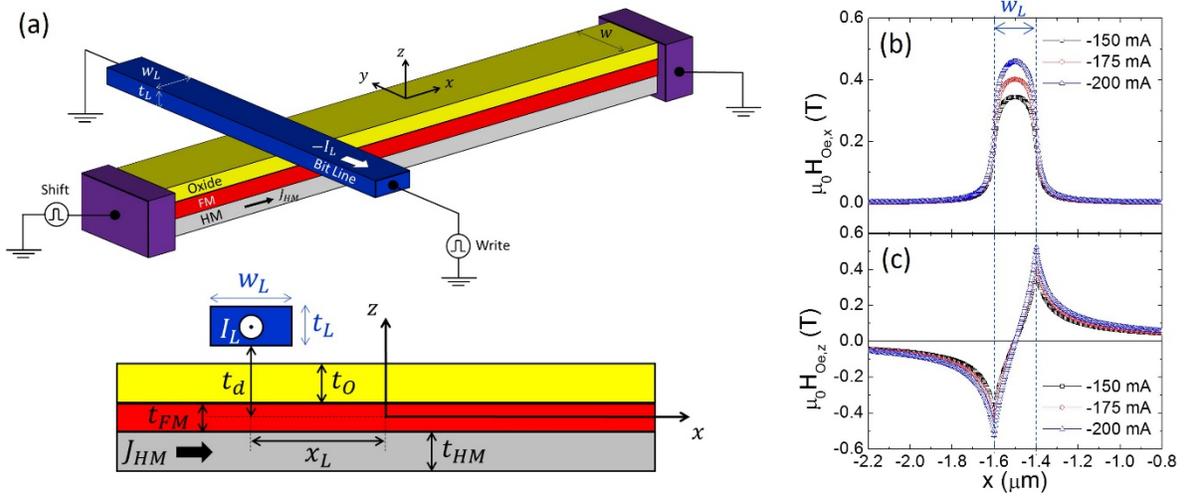

**Figure 1. Single bit line nucleation mechanism.** (**a**) Schematic of the HM/FM/oxide multilayer with a single orthogonal bit line along which a current pulse ($I_L$) is injected to generate the Oersted field ($\vec{B}_{Oe}$) needed to nucleate a pair of DWs in the FM layer. The centre of the bit line is at $(x_L, z_L) = (-1.5 \text{ μm}, t_d + \frac{t_L}{2})$ from the centre of the FM layer (0,0), where $t_d = 3$ nm represents the vertical distance between the bottom edge of the bit line and the centre of the FM layer ($x = 0$). The cross section of the bit line is $w_L \times t_L = 200 \times 50$ nm$^2$. (**b**)-(**c**) Longitudinal ($B_{Oe,x}$) and perpendicular ($B_{Oe,z}$) components of the Oersted field $\vec{B}_{Oe} = \mu_0 \vec{H}_{Oe}$ plotted against the *x* position along the FM strip for three different currents: $I_L = -150$ mA, $-175$ mA and $-200$ mA.

To illustrate the strong influence of the out-of-plane component of the Oersted field $B_{Oe,z}$, we first evaluated the DW nucleation process for systems schematically depicted in Fig. 1, where the nucleation is achieved by injecting a current pulse $I_L$ along the conventional single bit line. A *down* domain separated by an *up/down* DW and a *down/up* DW is initially located approximately 800 nm from the centre of the bit line (see Fig. 2). To nucleate a new *down* domain, a current pulse ($I_L(t)$) with the duration of 0.5 ns is injected along one bit line. The geometry and dimensions are identical to those shown in Fig. 1. Micromagnetic simulations indicate that a minimum amplitude of $I_L \approx 200$ mA is needed to promote the local magnetization reversal under the bit line (no reversal was achieved for $I_L \leq 175$ mA).

Transient snapshots during and after the current pulse are shown in Fig. 2 and indicate that the initial *down* domain is drastically perturbed by the Oersted field $B_{Oe,z}$; indeed, $B_{Oe,z} \approx$ 50 mT at the location of the initial *down* domain (see Fig. 1(c)). Such a high positive perpendicular field pushes the DWs at both sides of the *down* domain towards each other and is sufficiently strong to annihilate them (see snapshots at 0.5 ns and 0.7 ns in Fig. 2). Consequently, the DW nucleation by a single bit line is not an appropriate method for writing information in DW-based spintronic devices; thus, other procedures need to be developed to write highly packed information without disturbing the data already present in the FM strip.

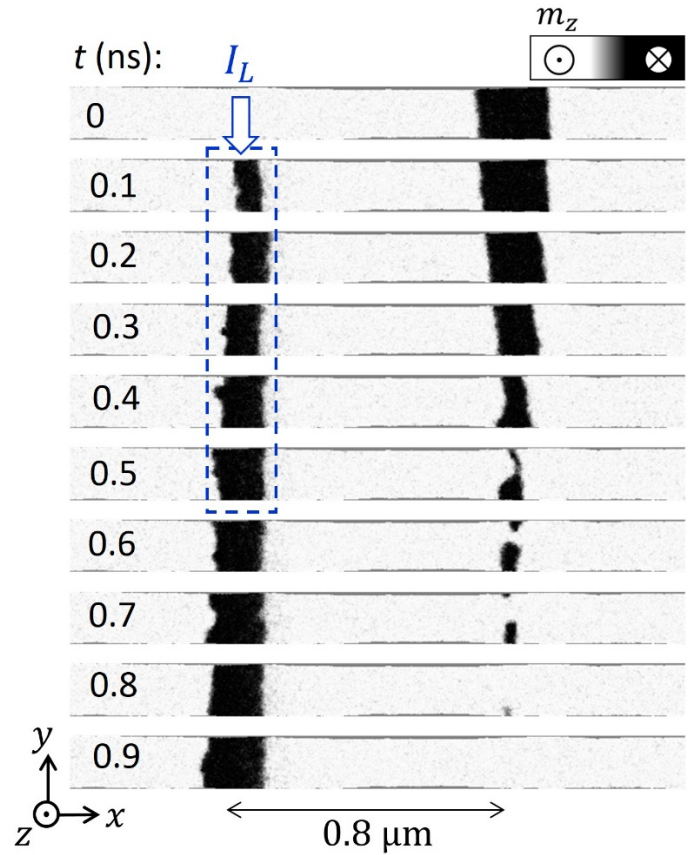

**Figure 2. Influence of the Oersted field generated by a single bit line on existing domain walls in a ferromagnetic strip.** Micromagnetic snapshots during the DW nucleation of a *down* domain under the injection of a short current pulse through a single bit line. Initially, ($t = 0$), the other reversed domain is placed approximately 800 nm from the nucleation point. The amplitude of the bit line pulse is $I_L = 200$ mA, and it is applied between $0 < t \leq 0.5$ ns. All dimensions are identical to those in Fig. 1. The Oersted field generated by the bit line pulse is sufficiently strong to annihilate the initial *down* domain.

**DW nucleation by a double bit line under current pulses along the HM**

As we have shown in the previous section, the main drawback of using a single bit line to nucleate DWs is the unwanted high out-of-plane component of the Oersted field $B_{Oe,z}$ far away from the nucleation point. To overcome this limitation, here we propose the use of a double bit line to generate a localized longitudinal Oersted field ($B_{Oe,x}$) with a negligible out-of-plane component ($B_{Oe,z} \approx 0$). This idea is based on the symmetry of the effective field due to the spin-orbit torques when an electrical current is injected through the HM (see the discussion below).

A schematic of this method is shown in Fig. 3(a), where the current pulse ($I_L$) flows in opposite directions along both bit lines: along the $+y$ direction in the bottom bit line ($+I_L$) and along the $-y$ direction in the top bit line ($-I_L$). If these bit lines are symmetrically placed below and above the FM layer ($t_d^T = t_d^B = 3$ nm), respectively, the perpendicular component of the Oersted field $B_{Oe,z}$ is zero over the FM layer (Fig. 3(c)); it therefore does not perturb the magnetic state of other DWs already present (see Supplementary Information, note S1 for the marginal influence of the asymmetric case). Moreover, with the double bit line configuration, the longitudinal component $B_{Oe,x}$ is enhanced in the region between the bit lines; therefore, a strong and localized in-plane field can be confined in the FM layer. However, this longitudinal field cannot, by itself, promote the localized reversal of the out-of-plane magnetization; therefore, it must be supported by another mechanism. The strong localized Oersted field $B_{Oe,x}$ is intended to promote the longitudinal component of the magnetization ($m_x \approx \pm 1$) between the lines: the spin-orbit torque from the current injection through the heavy metal ($J_{HM}$) will drive the local out-of-plane reversal, as discussed below.

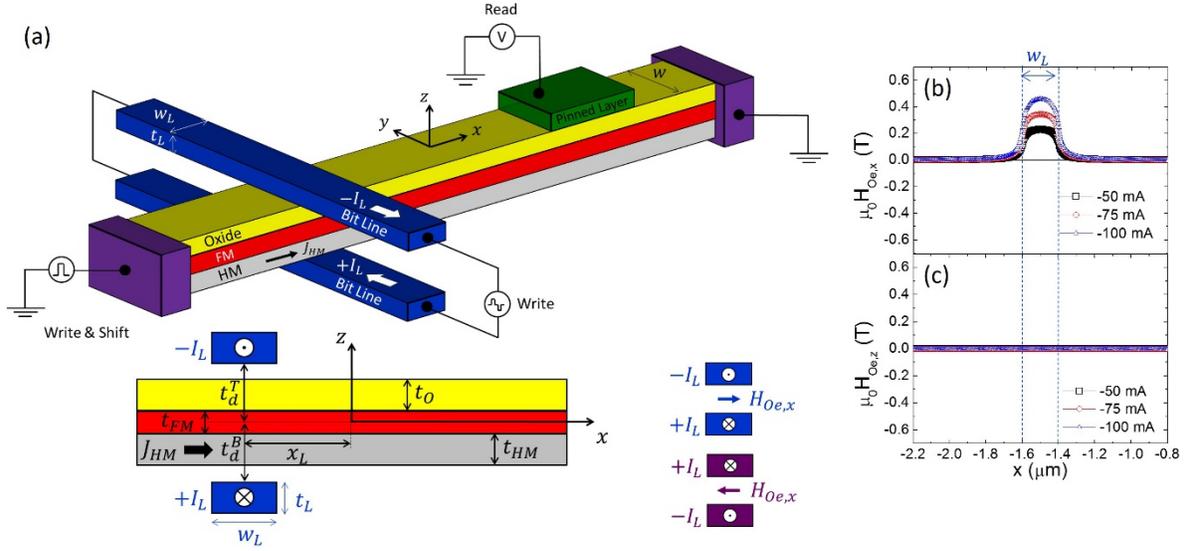

**Figure 3. Scheme with the double bit line to assist the current-driven DW nucleation. (a)** Schematic of the HM/FM/oxide multilayer with the two bit lines, where the current ($I_L$) flows along the $+y$ direction in the bottom bit line ($+I_L$) and returns along the $-y$ direction in the top bit line ($-I_L$). The centres of the bit lines are located at $(x_L, \pm z_L) = \left(-1.5\ \mu m, \pm \left(t_d + \frac{t_L}{2}\right)\right)$ with respect to the centre of the FM layer (0,0), where $t_d = t_d^T = t_d^B = 3$ nm is defined in the bottom graph of panel (a). The cross section of the bit lines is $w_L \times t_L = 200 \times 50\ nm^2$. **(b)-(c)** Longitudinal ($B_{Oe,x}$) and perpendicular ($B_{Oe,z}$) components of the Oersted field ($\vec{B}_{Oe} = \mu_0 \vec{H}_{Oe}$) generated by the double bit line plotted against the $x$ position along the FM strip for three different currents: $I_L = 50$ mA, 75 mA and 100 mA.

The effective field from the SL-SOT caused by the SHE ($\vec{H}_{SL}$) is now well established as being the main driving force of chiral DWs in asymmetric HM/FM stacks with high PMA[7,8,9,11,15]. This SL effective field is given by $\vec{H}_{SL} = H_{SL}^0 (\vec{m} \times \vec{\sigma})$, where $H_{SL}^0 = \frac{\hbar \theta_{SH} J_{HM}}{2|e|\mu_0 M_s t_{FM}}$ (Ref. 7). Here, $\hbar$ is Planck's constant, $|e|$ is the electric charge, $\mu_0$ is the permeability of free space, $t_{FM}$ is the thickness of the FM layer, and $\theta_{SH}$ is the spin Hall angle that determines the spin current/electric current ratio, $\theta_{SH} = J_s/J_{HM}$ [12,13,14]. $J_{HM}$ is the magnitude of the density current flowing through the HM ($\vec{J}_{HM}(t) = J_{HM}(t)\vec{u}_x$) and $\vec{\sigma} = \vec{u}_z \times \vec{u}_x$ is the unit vector of the spin current ($J_s$) generated by the SHE in the HM, which is orthogonal to both the direction of the electric current ($\vec{u}_x$) and the perpendicular direction ($\vec{u}_z$). For a general magnetization ($\vec{m} = m_x\vec{u}_x + m_y\vec{u}_y + m_z\vec{u}_z$), the SL effective field $\vec{H}_{SL}$ has both longitudinal $H_{SL,x}$ and perpendicular $H_{SL,z}$ components, $\vec{H}_{SL} = -H_{SL}^0 m_z \vec{u}_x + H_{SL}^0 m_x \vec{u}_z$, which are proportional to the perpendicular ($m_z$) and longitudinal ($m_x$) magnetization components, respectively. In high PMA systems, $\vec{m} \approx m_z \vec{u}_z$ with $m_x =$ $m_y = 0$ and $m_z \approx \pm 1$. However, if the magnetization is locally tilted towards the longitudinal direction ($m_x \gtrless 0$), a finite perpendicular component of the SL field appears: $H_{SL,z} = H_{SL}^0 m_x$. For a given polarity of $J_{HM}$, and depending on the sign of $m_x$, this field can be parallel or antiparallel to the out-of-plane component ($m_z$). Therefore, if sufficiently strong, $H_{SL,z}$ can promote the local out-of-plane switching when it is antiparallel to $m_z$. This SL-SOT symmetry, along with the possibility of generating a localized longitudinal field by injecting current pulses along the double bit line configuration, suggests a simple manner to nucleate DWs at will. Considering state-of-the-art material parameters[7,8,9,13,14] ($M_s = 600$ kA/m, $\theta_{SH} = 0.12$, $t_{FM} = 0.8$ nm) and typically injected current pulses through the HM ($J_{HM} = 2.5$ TA/m$^2$), the localized out-of-plane SL effective field reaches $\mu_0 |H_{SL,z}| \approx 300$ mT, a magnitude sufficient to nucleate domains and domain walls in high PMA multilayers.

**Discussion**

**Working principle and proof-of-concept**

The current-induced switching of an HM/FM/oxide stack under a global in-plane field has been experimentally demonstrated[7,12,13,14,15] and theoretically described by several works[30]. Here, we use the same physics to locally nucleate DWs under a local longitudinal field. The nucleation procedure is schematically explained in Fig. 4. Starting from a uniform out-of-plane magnetic state, either *up* ($m_z \approx +1$) or *down* ($m_z \approx -1$), current pulses are injected in the double bit line: $\pm I_L$ or $\mp I_L$, where $\pm$ ($\mp$) represent positive (negative) current along the top bit line, and negative (positive) current along the bottom one. The generated Oersted field ($H_{Oe,x}$) promotes the longitudinal component of the local magnetization ($m_x \approx \pm 1$) in the FM area between the bit lines. Its direction, along $m_x \approx +1$ or $m_x \approx -1$, is selected by the current direction along the double bit lines: $m_x \approx -1$ for $\pm I_L$ ($H_{Oe,x} < 0$) or $m_x \approx +1$ for $\mp I_L$ ($H_{Oe,x} > 0$). If a current pulse is simultaneously injected through the HM ($J_{HM}$), an out-of-plane SL effective field $H_{SL,z} = +H_{SL}^0 m_x$ ($\theta_{SH} > 0$) is generated between the lines. Depending on its magnitude and direction, it can locally switch the out-of-plane magnetization ($m_z$), resulting in the nucleation of a pair of DWs (see Fig. 4). Moreover, once nucleated, these DWs are displaced along the FM strip by the effective field $\vec{H}_{SL}$ itself[7,8].

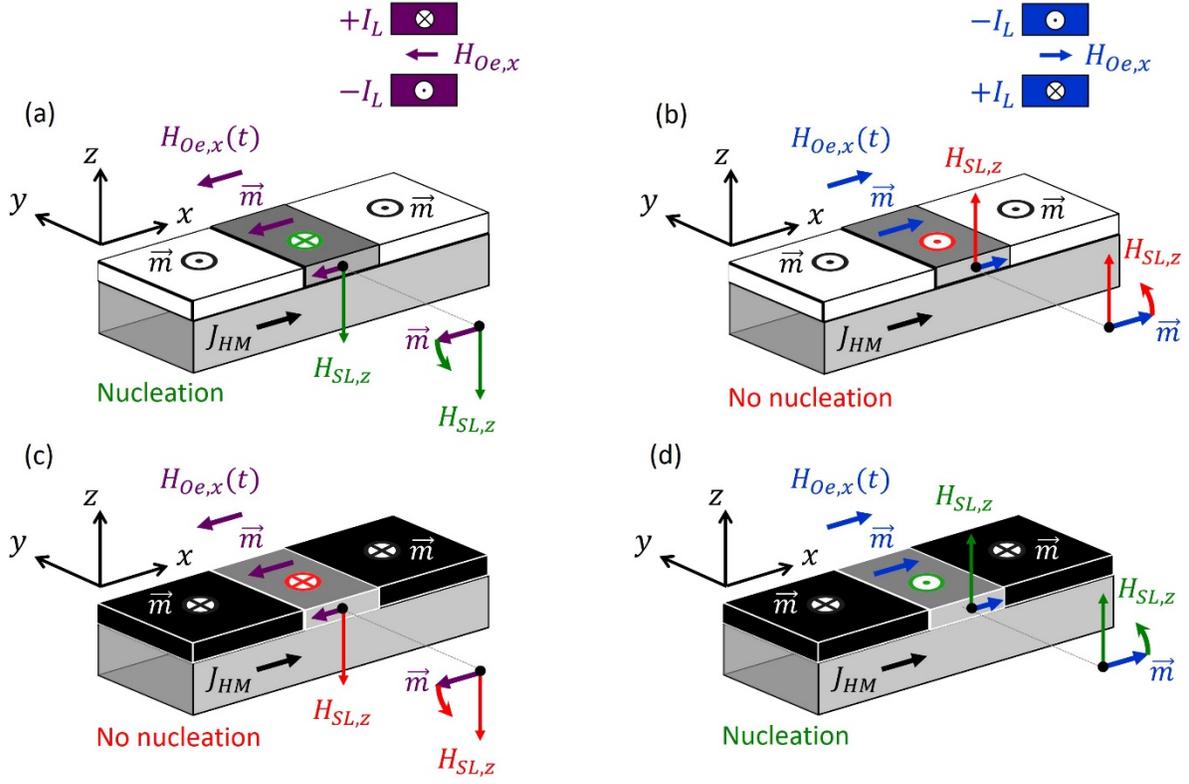

**Figure 4. Schematic of the operation mode based on the explained working principle. (a)** The initial state of the FM layer is magnetized *up* ($m_z = +1$). A local negative longitudinal field $H_{Oe,x}$ results in the FM by the current injection through the double bit line: $+I_L$ in the top bit line and $-I_L$ along the bottom one ($\pm I_L$). The grey region indicates the area of the FM layer between the bit lines, where the local magnetization points along the Oersted field generated by the double bit line ($m_x < 0$). At the same time, an electrical current is injected along the HM (black arrow), and the corresponding SL effective field $H_{SL,z}$ points along the negative *z*-axis ($H_{SL,z} < 0$, green arrow), locally reversing the initial magnetization from *up* ($m_z = +1$) to *down* ($m_z = -1$). A DW is nucleated and driven by the effective field itself. **(b)** Same as in (a), except that the local magnetization between the bit lines points along the positive *x*-axis, as it results from switching the current directions along the bit lines ($\mp I_L$). In this case, no DW nucleation is achieved because $H_{SL,z} > 0$ supports the initial magnetic state ($m_z = +1$). **(c)** and **(d)** correspond to (a) and (b) except that the initial state is magnetized *down*. The same physics as in (a) and (b) is straightforwardly understood.

To demonstrate the aforementioned spin-orbit nucleation procedure described in Fig. 4, micromagnetic simulations (see the Methods section) were performed under the simultaneous action of the current pulses through the double bit line ($\pm I_L$ or $\mp I_L$) and current pulses flowing through the HM ($J_{HM}$). Current pulses through the double bit line were applied with fixed magnitude ($\pm I_L = \pm 100$ mA or $\mp I_L = \mp 100$ mA). The duration of each $I_L$ pulse is 0.5 ns. The generated Oersted field is $H_{Oe,x} > 0$ for $\mp I_L$, and $H_{Oe,x} < 0$ for $\pm I_L$. Therefore, the magnetization is locally forced to point along $+x$ ($-x$) for $\mp I_L$ ($\pm I_L$). Simultaneously, a train of unipolar current pulses is injected through the HM with $J_{HM} = 2.5 \text{ TA/m}^2$. The duration of these pulses is also 0.5 ns, and the pulses are synchronized in time with the pulses along the double bit line. The time between consecutive pulses is also fixed to 0.5 ns. With the aim of evaluating the writing and shifting mechanism, we recorded the temporal evolution of the out-of-plane component of the magnetization $m_z(t,x)$ at different locations along the FM strip: point A is located between the bit lines ($x_A = -x_L$), whereas point B is located at $x_B = +540$ nm from the centre of the FM layer ($x = 0$). This point B would be the location of a reading magnetic tunnel junction, where the information would be read in the form of variations of the electrical resistance as the domains are shifted along the FM strip. As a proof-of-concept, simulations were intended to write 8 bits of information, where the binary elements "0" and "1" are represented by domains magnetized *up* ($m_z \approx +1$) and *down* ($m_z \approx -1$) respectively. The first example was implemented to write and shift along the FM layer a sequence of bits '10101010'; therefore, a train of eight bipolar current pulses was injected along the double bit line. The resulting longitudinal component of the Oersted field (at $x_A$) between the bit lines ($B_x(x_A, t)$) is shown in red colour on the top graph of Fig. 5(a). The unipolar current pulses in the HM are also shown in the same graph (black curves). The temporal evolution of the out-of-plane component of the magnetization $m_z(x, t)$ at points A ($x_A$) and B ($x_B$) are depicted in the bottom graph of Fig. 5(a). The corresponding snapshots of the magnetization along the FM strip are shown in the right panel of Fig. 5(a). As shown, $m_z(t)$ at A closely follows the evolution of the longitudinal Oersted field $B_x(x_A, t)$, indicating that the nucleation of the *up* and *down* domains is efficiently controlled by the proposed mechanism. Moreover, the coded information is shifted by the current flowing along the HM ($J_{HM}$) and the complete sequence of bits remains essentially unperturbed when it reaches point B, where the written information can be read in the form of variations of the electrical resistance. Note that the size of the initially nucleated domain is imposed by the double bit line width $w_L = 200$ nm, whereas the size of the shifted domains is determined by the DW velocity ($v_{DW} \sim J_{HM}$) and the pulse length along the HM ($\tau = 0.5 \text{ } ns$). Fig. 5(b) shows another example of the writing and shifting for a different sequence of bits: '11100100', demonstrating the robustness of the proposed method. Movies of these nucleation and shifting processes are included in the Supplementary Information.

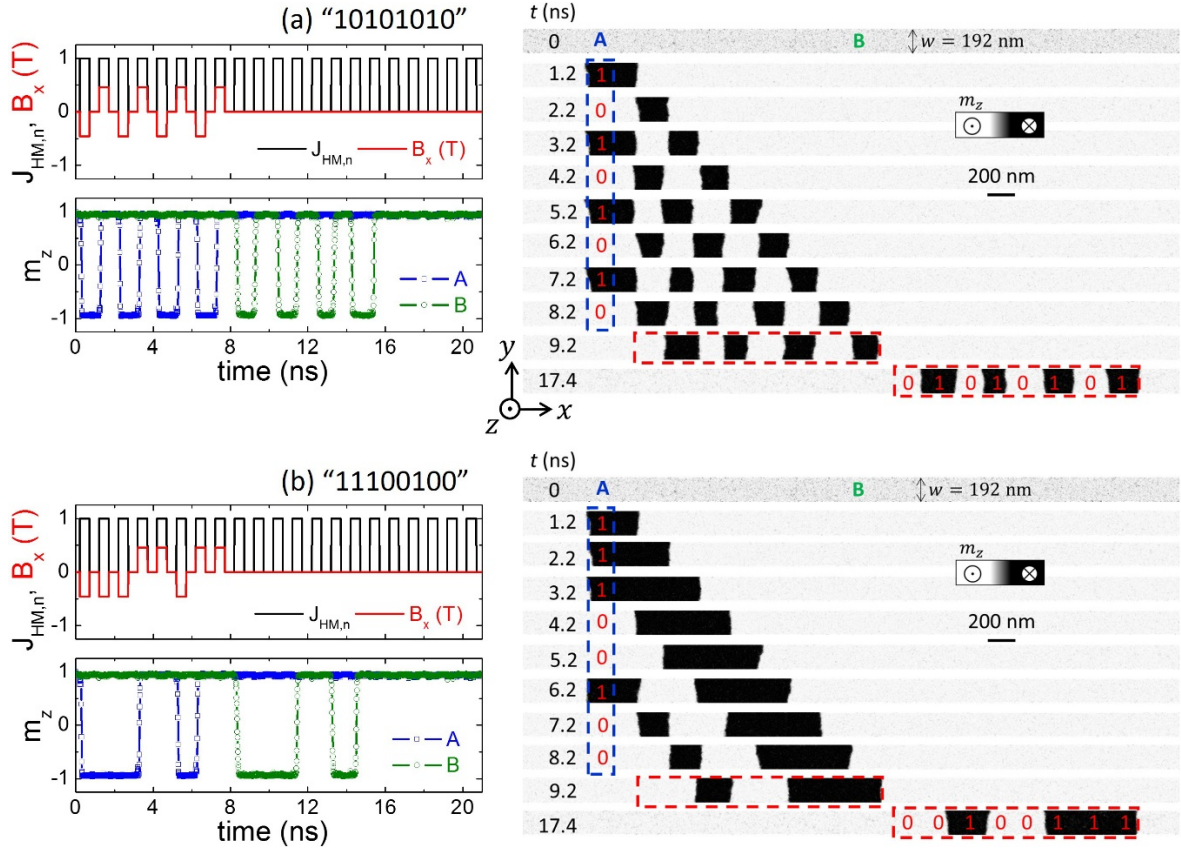

**Figure 5. Proof-of-concept. (a)** Temporal evolution of the local longitudinal magnetic field ($B_{Oe,x} \equiv B_x$) at the double bit line location (top-left graph, red lines). The sequence of current pulses through the double bit line is intended to write a '10101010' sequence of bits. The magnitude of the current through the double bit line is $I_L = 100$ mA. The dimensions of the bit lines are those given in Fig. 3, and the FM width is $w = 192$ nm. The amplitude of the unipolar current pulses along the HM is $J_{HM} = 2.5$ TA/m$^2$. Black lines in the top-left graph depict the temporal evolution of this current, normalized by a factor of 2.5 TA/m$^2$, $J_{HM,n} = J_{HM}/(2.5 \text{ TA/m}^2)$. The temporal evolution the out-of-plane magnetization $m_z$ is presented in the left-bottom graphs at two different points: A (blue) and B (green), as marked in the first snapshot of the right column. Point A is located between the bit lines, $x_A = -x_L$, whereas point B is located at the centre of the FM layer $x_B = +540$ nm from the centre of the FM layer ($x = 0$). The snapshots at the right panel show the out-of-plane magnetization ($m_z$) at different instants. The dashed blue line indicates the location of the double bit line, whereas the dashed red rectangle depicts the bit sequence just after the end of the writing pulse sequence ($t = 9.2$ ns) and when all bits have passed through point B ($t = 17.4$ ns). **(b)** Same as in (a) for a different bit sequence through the double bit line: '11100100'.

To show the robustness of the proposed nucleation mechanism, we have repeated the previous analysis for other sets of material parameters with PMA ($K_{eff} = K_u - \frac{1}{2}\mu_0 M_s^2 > 0$). The results are shown in Fig. 6. We first fixed $M_s = 0.6$ MA/m and varied the PMA constant $K_u$. The critical current along the bit lines ($I_{L,c}$) as a function of $K_u$ is depicted in Fig. 6(a) under 0.5 ns-long current pulses of different magnitudes ($J_{HM}$). The values of $J_{HM}$ were chosen well above the propagation threshold (approximately 0.3 TA/m$^2$) because of the defects (see the Supplementary Information, note S2). In Fig. 6(b), $K_u$ is fixed to $K_u = 0.6$ MJ/m$^3$, and the saturation magnetization $M_s$ is varied. $I_{L,c}$ increases monotonously with $K_u$ (Fig. 6(a)) and decreases monotonously as $M_s$ is increased (Fig. 6(b)). The same trends are observed in Figs. 6(c) and 6(d) for other parameter sets (0.9 MJ/m$^3 \leq K_u \leq 1.5$ MJ/m$^3$, and 0.7 MA/m $\leq M_s \leq 1.25$ MA/m). Similar results were also obtained for other values of the DMI parameter in the range $0.5 \frac{\text{mJ}}{\text{m}^2} \leq D \leq 1.5 \frac{\text{mJ}}{\text{m}^2}$ (not shown). Therefore, the nucleation mechanism is robust for a wide range of material parameters with PMA.

Additionally, we can estimate the magnitude of the current along the bit lines ($I_{L,c}$) needed to nucleate DWs, describing the micromagnetic results in Fig. 6. The effective perpendicular field given by $\mu_0 H_{eff} = \frac{2K_{eff}}{M_s}$ with $K_{eff} \approx K_u - \frac{1}{2}\mu_0 M_s^2$, represents the magnitude of the longitudinal field $\mu_0 H_x$ required to saturate the out-of-plane magnetization along the longitudinal field (*i.e.,* along the $x$-axis). In our nucleation scheme, this longitudinal field is generated by the current pulse along the double bit line. Assuming the double bit lines are very long along the $x$-axis, this field can be estimated from Ampere's law[29] as $\mu_0 H_x \approx \mu_0 \frac{I_L}{w_L}$. Therefore, the minimum value of $I_L$ for the nucleation is $I_L \geq I_{L,c} \approx \frac{2w_L}{M_s} K_{eff}$ which, qualitatively matching the trends observed in Fig. 6 (see solid lines in Fig. 6, Approx.). The $I_{L,c}$ increases linearly with increasing $K_u$ for a fixed $M_s$, whereas it monotonically decreases as $M_s$ decreases for a fixed $K_u$. This approximation overestimates $I_{L,c}$ with respect to full micromagnetic calculations. However, this overestimation is expected because the approximation neglects the action of the out-of-plane component of the SHE effective field ($H_{SL,z} = H_{SL}^0 m_x$, with $H_{SL}^0 = \frac{\hbar \theta_{SH} J_{HM}}{2|e|\mu_0 M_s t_{FM}}$), promoting the local magnetization reversal as soon as a finite longitudinal component of the magnetization arises ($m_x \neq 0$). The presented micromagnetic results also indicate that $I_{L,c}$ decreases as $J_{HM}$ increases.

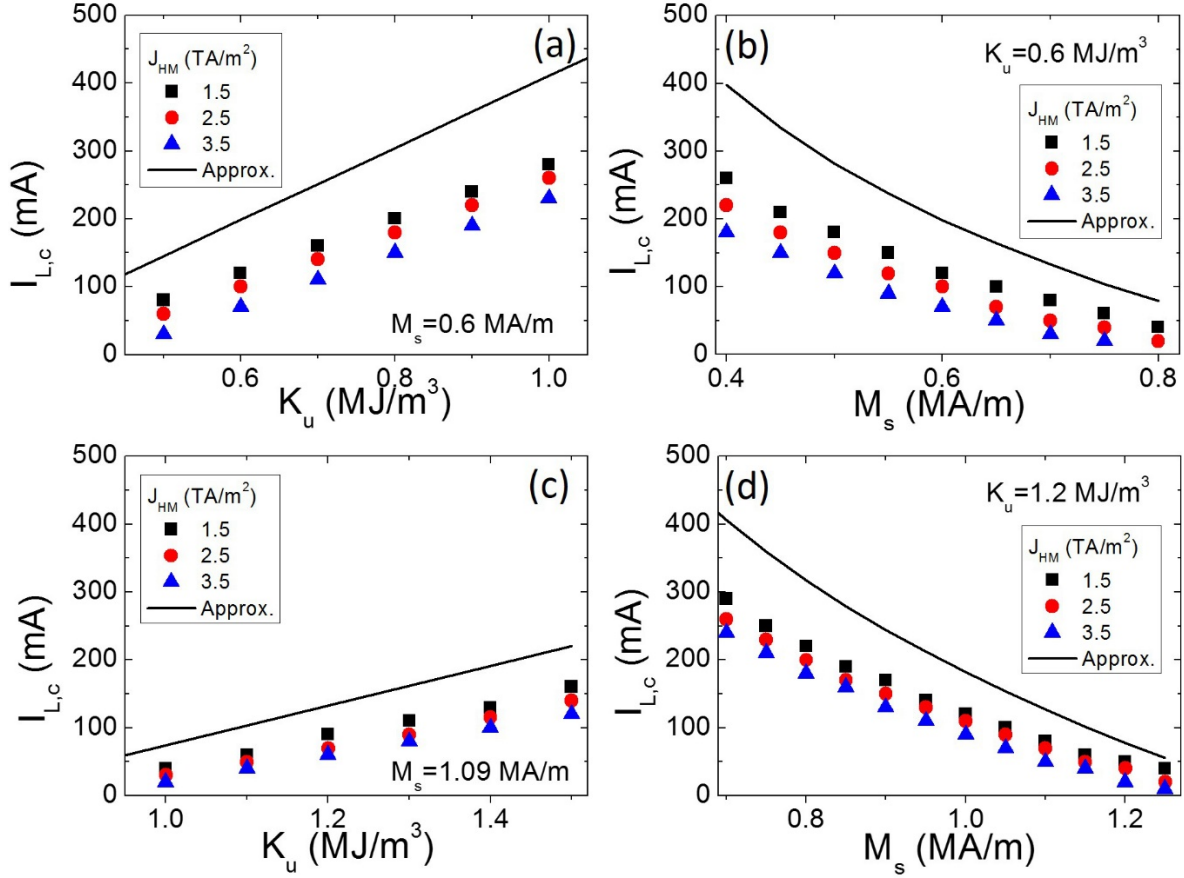

**Figure 6. Critical current along the double bit line ($I_{L,c}$) to achieve DW nucleation and shifting as a function of the material parameters.** (a) $M_s = 0.6$ MA/m. (b) $K_u = 0.6$ MJ/m$^3$. The other material parameters are the same as those in Fig. (5). (c) $I_{L,c}$ vs $K_u$ for $M_s = 1.09$ MA/m. (d) $I_{L,c}$ vs $M_s$ for $K_u = 1.2$ MJ/m$^3$. Solid lines correspond to the estimation for $I_{L,c}$ described in the text.

**Conclusions**

We have reported a general method to locally control the current-induced magnetization reversal assisted by a local longitudinal field within ferromagnetic strips with high PMA. The local longitudinal Oersted field is generated by a double bit line orthogonal to the ferromagnetic layer, where short current pulses flow along opposite directions. This field aligns the local magnetization along the longitudinal strip axis, and its direction, either $+x$ or $-x$, is manipulated by changing the directions of the current polarity in the bit lines. The local reversal of the out-of-plane direction of the magnetization is promoted by the perpendicularly effective field due to the spin Hall effect in the heavy metals, resulting in the efficient nucleation of domain walls, which are also driven along the ferromagnetic strip due to the spin Hall effective field itself. This simple procedure based on the symmetry of the spin orbit torque is fast, versatile and easy to implement. From a technical viewpoint, our method requires fabricating an injection double bit line such that the bottom bit line is electrically isolated from the heavy metal under the orthogonal ferromagnetic layer. This fabrication can be achieved by placing the bottom injection bit line under the heavy metal or by burying it within the heavy metal with an insulating covering.

Although this lithography step may not be straightforward, it should be accessible through current state-of-the-art fabrication methods. The multilayers and the material parameters explored in our numerical simulations are only an example of the efficient and rapid nucleation mechanism, which could be further optimized to reduce the energy consumption and to improve the recording density and the shifting time of stored information in racetrack memory DW-based devices.

**Methods**

**Oersted field generated by the single and the double bit line configurations**

The Oersted field of a current flowing in the single bit line was initially computed numerically by solving the Poisson equation for the potential vector using the finite element method. The results are also analytically described by directly solving Biot-Savart's law[29]. The longitudinal $B_{Oe,x}$ and perpendicular $B_{Oe,z}$ components of this Oersted field are expressed as

$$B_{Oe,x}(X,Z) = \frac{\mu_0 J_L}{2\pi} \left[ F\left(X + \frac{w_L}{2}, Z + t_L\right) - F\left(X + \frac{w_L}{2}, Z\right) + F\left(-X + \frac{w_L}{2}, Z + t_L\right) - F\left(-X + \frac{w_L}{2}, Z\right) \right] \quad (1)$$

$$B_{Oe,z}(X,Z) = \frac{\mu_0 J_L}{2\pi} \left[ G\left(X + \frac{w_L}{2}, Z + t_L\right) - G\left(X + \frac{w_L}{2}, Z\right) - G\left(-X + \frac{w_L}{2}, Z + t_L\right) + G\left(-X + \frac{w_L}{2}, Z\right) \right] \quad (2)$$

where $J_L = I_L/(w_L t_L)$ is the current density along the bit line, the coordinate $X = x - x_L$ represents the longitudinal distance from the centre of the bit line, and $Z = z - t_d$ represents the vertical distance from the bottom edge of the bit line (see Fig. 1(a) and Fig. 3(a) for definitions of $x_L$ and $t_d$). Functions $F(x,z)$ and $G(x,z)$ are defined as

$$F(x,z) = z \arctan\left(\frac{x}{z}\right) + \frac{1}{2}x \log(x^2 + z^2) \tag{3}$$

$$G(x,z) = -z + x \arctan\left(\frac{z}{x}\right) + x \log(x^2 + z^2) \tag{4}$$

For the double bit line configuration, where the current flows along opposite directions through both lines, the components of the generated Oersted field in the FM layer can be deduced from the former expressions by using symmetry arguments and the superposition principle.

**Micromagnetic details**

The current-induced DW dynamics is governed by the Gilbert equation augmented by the Slonczewski-like spin-orbit torque (SL-SOT) [7,11,28,30,31,32],

$$\frac{d\vec{m}}{dt} = -\gamma_0 \vec{m} \times (\vec{H}_{eff} + \vec{H}_{th}) + \alpha \vec{m} \times \frac{d\vec{m}}{dt} + \vec{\tau}_{SL} \tag{5}$$

where $\gamma_0$, $\alpha$ and $\vec{m}(\vec{r},t) = M(\vec{r},t)/M_s$ denote the gyromagnetic ratio, the Gilbert damping constant and the normalized local magnetization to the saturation value ($M_s$), respectively. $\vec{H}_{eff}$ is the deterministic effective field that includes exchange, magnetostatic interactions, the Oersted field generated by the bit lines, and the uniaxial anisotropy and the Dzyaloshinskii-Moriya interaction (DMI). $\vec{H}_{th}$ is the stochastic thermal field[33]. The last term, $\vec{\tau}_{SL}$, is the SL-SOT due to the spin Hall effect (SHE),

$$\vec{\tau}_{SL} = -\gamma_0 \frac{\hbar \theta_{SH} J_{HM}}{2\mu_0 |e| M_s t_{FM}} \vec{m} \times (\vec{m} \times \vec{\sigma}) = -\gamma_0 \vec{m} \times \vec{H}_{SL} \tag{6}$$

where $\vec{H}_{SL} = H_{SL}^0 (\vec{m} \times \vec{\sigma})$ with $H_{SL}^0 = \frac{\hbar \theta_{SH} J_{HM}}{2\mu_0 |e| M_s t_{FM}}$. Here, $\hbar$ is Planck's constant, $|e|$ is the electric charge, $\mu_0$ is the permeability of free space, $t_{FM}$ is the thickness of the ferromagnetic layer, and $\theta_{SH}$ is the spin Hall angle. $J_{HM}$ is the magnitude of the density current along the HM layer ($\vec{J}_{HM} = J_{HM}(t)\vec{u}_J$), and $\vec{\sigma} = \vec{u}_z \times \vec{u}_J$ is the unit vector of the spin current generated

by the SHE in the HM, which is orthogonal to both the direction of the electric current ($\vec{u}_J = \vec{u}_x$) and the perpendicular direction ($\vec{u}_z$). Typical parameters for a HM/FM/oxide multilayer with strong DMI are considered for the results presented in the main text[7,8]. Unless otherwise indicated, these parameters are $M_s = 0.6 \times 10^6$ A/m, $A = 2.0 \times 10^{-11}$ J/m, $K_u = 0.6 \times 10^6$ J/m$^3$, $D = -0.75$ mJ/m$^2$, $\alpha = 0.1$ and $\theta_{SH} = 0.12$. The thickness of the FM layer is $t_{FM} = 0.8$ nm. The length of the FM strip is $\ell = 6134$ nm, and different widths ($w = 96$ nm and $w = 192$ nm) were evaluated, which led to similar results. Except when the opposite case is indicated, disorder and thermal effects at room temperature ($T = 300$ K) are taken into account to mimic realistic conditions. The local disorder and imperfections are taken into account in the simulations by considering both randomly generated roughness at the strip edges[33] and grains in the strip body[11]. A typical edge roughness with a characteristic length of 9 nm is considered in the present study. In addition, we assume that the easy-axis anisotropy direction is distributed among the length scale defined by the characteristic grain size of 10 nm. The direction of the uniaxial anisotropy of each grain is mainly directed along the perpendicular direction (z-axis) but with a small in-plane component randomly generated over the grains. The maximum percentage of the in-plane component of the uniaxial anisotropy unit vector is 10%. In this work, five different edge roughness and grain patterns were evaluated to confirm the validity of the method under realistic conditions. Micromagnetic simulations were performed using the Mumax micromagnetic simulation program[34]. The samples were discretized using a 2D grid of 3 nm wide cells. The time step used to solve the magnetization dynamics was $\Delta t = 0.1$ ps, and several tests were performed with smaller time steps to check that the validity of the presented results.

**Acknowledgements**

This work was supported by project WALL, FP7- PEOPLE-2013-ITN 608031 from the European Commission, project MAT2014- 52477-C5-4-P from the Spanish government, and project SA282U14 from the Junta de Castilla y Leon.

**Author Contributions**

E.M. and V.R. conceived and coordinated the project. V.R., O.A., L.S.-T., and E.M. performed the micromagnetic simulations. All authors analysed and interpreted the results. E.M. wrote the manuscript with the assistance of V.R and O.A. All authors commented on the manuscript.

**Additional Information**

Supplementary information accompanies this paper at http://www.nature.com/srep

Competing financial interests: The authors declare no competing financial interests.

**Data Availability Statement**

The datasets generated during and/or analysed during the current study are available from the corresponding author upon reasonable request.